\documentclass[preprint,eqsecnum,aps,showpacs]{revtex4}

\usepackage{amssymb}

\newcommand{\be}{\begin{equation}}
\newcommand{\ee}{\end{equation}}
\newcommand{\ba}{\begin{eqnarray}}
\newcommand{\ea}{\end{eqnarray}}
\newcommand{\ban}{\begin{eqnarray*}}
\newcommand{\ean}{\end{eqnarray*}}

\begin{document}


\title{A Continuum Description of Rarefied Gas Dynamics (I)---
Derivation From Kinetic Theory}
\author{Xinzhong Chen}
\affiliation{Astronomy Department, Columbia University, New York, NY 10027}
\author{Hongling Rao}
\affiliation{Microeletronics Sciences Laboratories, Columbia University,
New York, NY 10027}
\author{Edward A. Spiegel}
\affiliation{Astronomy Department, Columbia University, New York, NY 10027 }

\date{\today}

\begin{abstract}
We describe an asymptotic procedure for deriving continuum equations
from the kinetic theory of a simple gas.  As in the works of Hilbert, of
Chapman and of Enskog, we expand in the mean flight time of the
constituent particles of the gas, but we do not adopt the Chapman-Enskog
device of simplifying the formulae at each order by using results from
previous orders.  In this way, we are able to derive a new set of
fluid dynamical equations from kinetic theory, as we illustrate here for
the relaxation model for monatomic gases.  We obtain a
stress tensor that contains a dynamical pressure term (or bulk viscosity)
that is process-dependent and our heat current depends on the gradients of
both temperature and density.  On account of these features, the equations
apply to a greater range of Knudsen number (the ratio of mean free path
to macroscopic scale) than do the Navier-Stokes equations, as we see in
the accompanying paper.  In the limit of vanishing Knudsen number, our
equations reduce to the usual Navier-Stokes equations with no bulk
viscosity.
\end{abstract}


\maketitle


\section{Introduction}

The derivation of fluid equations from kinetic theory is often effected
by an expansion of the solution of the kinetic equation in the Knudsen
number, the ratio of the mean free path to the characteristic macroscopic
scale.  The method generally in use was developed by Chapman and
Enskog~\cite{cha61} whose aim in part was to remedy deficiencies in
the earlier work of Hilbert~\cite{gra63,caf83}.  Chapman and Enskog
derived the Euler equations in zeroth order and the Navier-Stokes
equations at first order, starting from the Boltzmann equation of kinetic
theory.  But the Navier-Stokes equations do not adequately describe the
dynamics of fluids when the Knudsen number is not very small.  The
inadequacies of the N-S equations for dealing with the problems of
rarefied media are well documented~\cite{uhl63} and we shall cite
empirical evidence for this when we compare our theory with experiment
in papers II and III of this series.  Problems arise for example in the
study of shock waves since the thickness of a shock
wave is generally of the order of the mean free path of the particles that
make up the medium through which the shock propagates (paper III).
However, the difficulty has little if anything to do with the
nonlinearity of the shock waves, as we may see from the similar failure
of the N-S equations to predict accurately the propagation of linear sound
waves when their periods are comparable to the mean flight times of the
constituent particles (paper II).

To improve matters, one may try solving the kinetic equations
directly~\cite{uhl63}, but this leads to a problem in higher dimensions
than is encountered in solving the continuum fluid equations.
Alternatively, one may seek to derive from the kinetic theory a set of
fluid equations with a greater domain of validity than the N-S equations.
This was the aim of attempts to go to higher order in the development of
the kinetic equation in Knudsen number.
Higher order approximations, such as the Burnett or super Burnett
developments~\cite{bur35a,bur35b} (and that of Woods~\cite{woo93})
have hardly improved matters and the resulting equations do not work well
when the Knudsen number is not infinitesimal~\cite{gra63}.  Moreover,
the continuum equations found in these higher approximations are very
complicated so that they do not seem to repay the effort involved in their
use.

Some hope for improvement of the situation was raised when
Grad~\cite{gra58} introduced his moment method, whose leading order
results are the Navier-Stokes equations.  However, the further development
of Grad's method~\cite{jou93,mul98} does not produce very
rapid convergence nor does it easily give very accurate solutions of the
problems mentioned here (as we shall see in paper II).
Levermore~\cite{lev96} has proposed an alternate moment method
that avoids the failures of causality that sometimes arise with
the Grad method.  The newer method has produced interesting results on
shock structure~\cite{lev98} but as yet has not given results with heat
flux.

Other approaches that are being explored are the resummation of the terms
of the Chapman-Enskog  expansion~\cite{ros89,ros93,sle97} and the
flux-limited diffusion theory~\cite{lev79}; both of these
approaches have proved useful in radiative transfer theory (the former in
\cite{unn66,che00b} and the latter in \cite{lev81}).  A promising and
interesting approach is by way of thermodynamic (or phenomenological)
models~\cite{jou93,mul98} that give good agreement with
experiment~\cite{jou93}.  The relativistic extensions of these
latter procedures~\cite{isr63} produce hyperbolic systems and do not
violate causality.

Here we propose a modification of the asymptotic procedure of
Chapman and Enskog.  We begin, as they do, with an expansion in mean free
paths of the constituent particles, {\it a la} Hilbert.  However, we
do not simplify the results at a given order by introducing expressions
(mainly for time derivatives of fluid quantities) from previous orders
as did Chapman and Enskog.  As we shall see, this
seemingly small difference in the methods makes for large differences
in the results when they are applied to cases with finite mean free
paths.  As mentioned in~\cite{che00a} and more fully documented in
paper II, even our first-order asymptotics provides acceptable
accuracy when comparison is made with experimental results.

In this paper, we derive continuum equations from the relaxation
model of kinetic theory~\cite{cer88,bha54,wel54} (also called the
BGK model). In the relaxation model, the approach toward
equilibrium takes place in a relaxation time, $\tau$,
which is determined by the mean flight time of the constituent
particles of the medium.  So the analogue of Hilbert's approach
is here an expansion in relaxation time.  In our version of this
development, we shall not recycle the lower orders through the current
order as is done in Chapment-Enskog theory.  The fluid equations
folowing from our procedure, generalize the Navier-Stokes equations,
which we shall recover from ours by a simple development for the
stress tensor
and the heat flux in terms of $\tau$, which is effectively
the Knudsen number.  In that sense, a byproduct of our development
is a relatively simple derivation of the Navier-Stokes equations,
one that avoids some of the complications of the Chapman-Enskog
method.

\section{Some Kinetic Theory}

Consider a gas made up of identical particles of mass
$m$ obeying classical Hamiltonian dynamics.  The phase space of a
single particle is six-dimensional and its coordinates are spatial
position, ${\bf x}$, and velocity, ${\bf v}$.  The expected number
of particles in a phase volume $d{\bf x} d{\bf v}$ is $f({\bf x},
{\bf v}, t)d{\bf x} d{\bf v}$ where $f$ is the density in phase
space (or one-particle distribution function). For a Hamiltonian
system the phase flow is incompressible in the sense that its
six dimensional velocity is solenoidal, and the relaxation model of
kinetic theory~\cite{bha54,kog69,wel54} that we adopt here is
\begin{equation}
\mathcal{D} f = {f_0-f\over \tau} \ ,  \label{x1}
\end{equation} with the following notation.  The streaming
operator is \begin{equation} \mathcal{D} = \partial_t + v^i
\partial_{x^i} + a^i \partial_{v^i} \label{CalD} \end{equation}
where $i=1,2,3$, repeated indices are summed,
$\partial_{x^i}=\partial/\partial x^i$,
$\partial_{v^i}=\partial/\partial v^i$, $a^i$ represents the acceleration
of a particle caused by an external force and $\tau$ is the time
scale on which the system relaxes to the equilibrium, $f_0$.

While we prefer to think of (\ref{x1}) as a model, it may be also
considered as an approximation to the Boltzmann equation where $\tau$
and $f_0$ are taken as approximations to functionals of $f$ that appear
in the Boltzmann collision operator.  With either interpretation,
the relaxation time, $\tau$, is of the order of a collision time.  Though
$\tau$ might in principle depend on ${\bf v}$, we here assume that it
depends on just the local values of the macroscopic fields.  To give an
expression for $\tau$ we then need to specify the macroscopic quantities.

The mass density of the fluid is defined as \begin{equation}
\rho({\bf x}, t) = \int m f d{\bf v}\; , \label{density} \end{equation}
where the integration is over all of velocity space.  We may also
introduce the particle number density $n=\rho/m$.  A second important
macroscopic  quantity is the mean, or drift, velocity, defined as
\begin{equation} {\bf u}({\bf x}, t) = {1\over \rho}
\int m{\bf v} f d{\bf v}\; . \label{velocity} \end{equation}
We may then introduce the peculiar velocity ${\bf c} = {\bf v - {\bf u}}$
and use it to define the temperature as \begin{equation}
T = {1\over 3R\rho}\int {\bf c}^2 f d{\bf v}\; , \label{temp}
\end{equation}
where $R=k/m$ is the gas constant and $k$ is the Boltzmann constant.
We assume throughout this work that $f$ goes rapidly to zero as $|{\bf
v}|\rightarrow \infty$, so that $f$-weighted integrals over velocity
space are finite and well defined.

The macroscopic quantities are all functions of ${\bf x}$ and $t$, and,
when we speak of a temperature, we use the notion of
local thermodynamic equilibrium, in which equilibrium formulae are used
to describe nonequilibrium conditions locally in space and time.
In the relaxation version of the kinetic equation, the interaction
term on the right of (\ref{x1}), representing collisions amongst the
particles, drives the system toward the equilibrium given by $f_0$, which
we take to be the local Maxwell-Boltzmann distribution in the frame
locally comoving with the matter, namely
\begin{equation}
f_0  = n (2\pi R T)^{-{3\over 2}}
\exp\left(-{c^2\over2RT}\right) \ .  \label{MB} \end{equation}
This distribution depends on position and time only through its dependence
on the local macroscopic quantities, $n,\ {\bf u}$ and $T$ and it is
moreover a local equilibrium solution of the Boltzmann equation.

To make the description work well, we try to arrange that the
fields on which $f_0$ depends are most nearly those of the real
flow.  This `osculating' property of the assumed equilibrium may
be imposed by matching conditions that are inherited from the
theory of the Boltzmann equation.  There the quantities that are
conserved in two body collisions are the sums of the masses, the
velocities and the energies of the of the two colliding particles.
We let $\psi_\alpha$ with $\alpha=0,1,...,4$ be these summations
invariants; that is $\psi_0=m$, $\psi_i=mv_i$ and $\psi_4={1\over
2}m{\bf v}^2$ with $i=1,2,3$.  In Boltzmann theory, this
conservation property results in the orthogonality of the
$\psi_\alpha$ with the collision term in the kinetic theory.  This
property is generally ascribed to kinetic models including
the relaxation model. Therefore when
we multiply the right side of equation by any of the $\psi_\alpha$
and integrate over velocity space, we get zero. This gives us the
matching conditions
\begin{equation} \int \psi_\alpha f_0 d{\bf v}\; = \int
\psi_\alpha f d{\bf v}\; , \label{match} \end{equation}
provided that we may assume that $\tau$ does not depend on the velocity.
This condition ensures that the true fluid fields are the ones
appearing in (\ref{MB}).

Finally, we may specify that $\tau$ should be of the order of the mean
flight time of the constituent particles, that is, the mean free path of
particles divided by the mean speed.  If the collision cross-section of
the particles is independent of velocity, then simple collision theory
gives the result that $\tau \propto {1 \over \rho\sqrt{T}}$, so that,
generally speaking, $\tau$ is generally a function of the macroscopic
variables.

\section{Equations of Fluid Dynamics}

When when we multiply (\ref{x1}) by $\tau \psi_\alpha$
and integrate over ${\bf v}$, the right side is zero because of the
matching condition.  If we add the assumption that $\tau$ does not depend
on ${\bf v}$ we obtain these macroscopic equations:
\begin{eqnarray}
&  & \partial_t \rho + \nabla\cdot(\rho{\bf u}) = 0 \label{cont}
\\ &  & \rho\left(\partial_t {\bf u} + {\bf u}\cdot\nabla {\bf
u}\right) +\nabla \cdot \mathbb{P} = \rho {\bf a} \label{mom}
\\ &  & {3\over 2}\rho R\left(\partial_t T + {\bf u}\cdot\nabla T \right)
+ \mathbb{P}:\nabla{\bf u} + \nabla \cdot {\bf Q} = 0 \ ,
\label{heat}
\end{eqnarray}
where the semicolon stands for a double dot product and we have
introduced the pressure tensor (or second moment)
\begin{equation}
\mathbb{P}  = \int m {\bf c} {\bf c} f d{\bf v} \label{ptensor}
\end{equation} and the heat flux vector (a third moment) given by
\begin{equation}
{\bf Q} = \int {1\over 2} m {\bf c^2} {\bf c} f d{\bf v}\ . \label{flux}
\end{equation}
In deriving these equations, we have performed integrations by parts
under the integrations over velocity space.  These give no boundary terms
if, as we assume, $f$ goes rapidly enough to zero as $|{\bf c}|$ gets
large. (Strictly speaking, the $\mathbb{P}$ in (\ref{heat}) should be
the transpose, but as the stress tensor is symmetric in this work, we need
not indicate this.)

These macroscopic equations are a formal consequence of the kinetic
equation, (\ref{x1}). For them to be useful, we must supply adequate
expressions for $\mathbb{P}$ and ${\bf Q}$.  The Navier-Stokes
forms of these higher moments, which were first derived
phenomenologically, were obtained for small mean free paths by the
Chapman-Enskog method from the Boltzmann equation.  The N-S equations have
also been extracted from (\ref{x1}) by this method~\cite{cer88}.  Those
developments follow on the work of Hilbert, who introduced a series
expansion in mean free path for $f$ into the Boltzmann equation
\cite{caf83,cer88,gra63}.  In the case of the relaxation model, matters
are simpler because it has an essentially linear form and this permits a
clearer statement of the underlying approach to deriving the fluid
equations.  The procedure is to develop in $\tau$, on the
presumption that it is small compared to the macroscopic times that arise.
Such an asymptotic development for the case of the relaxation model begins
with an expansion of the form \begin{equation}
f = \sum_{m=0}^{\infty} f_{(m)} {\tau}^m  \ . \label{Hilb}
\end{equation}

When we substitute (\ref{Hilb}) into (\ref{x1}) we obtain a series of
equations for the $f_{(m)}$. The first of these is the anticipated
condition $f_{(0)}=f_0$.  The second approximation is simply
\begin{equation}
f_{(1)} = -\mathcal{D} f_0 \ .  \label{ser} \end{equation}
When we introduce expression (\ref{MB}) into this result we obtain
the more explicit relation \begin{equation}
f_{(1)} =-f_0\left[\mathcal{D}\ln \rho +\left({c^2\over 2RT}-\frac
{3}{2}\right)
\mathcal{D}\ln T +\frac{1}{RT} {\bf c}\cdot \mathcal{D}{\bf c}\right].
\label{x27}\end{equation}

Matters are quite simple up to this point but when we go to higher
orders, we encounter terms arising from the derivatives of $\tau$.
So we stop the expansion at this point since we already have enough in
first order to obtain an interesting generalization of the equations of
fluid mechanics.

To use the asymptotic results for $f$ in (\ref{ptensor})
and (\ref{flux}), we note that the higher moments can be written as
\begin{equation}
\mathbb{P} = \mathbb{P}_{(0)} + \tau \mathbb{P}_{(1)} + ...
\label{bbPn} \end{equation} and \begin{equation} {\bf Q} =
\mathbf{Q}_{(0)} + \tau \mathbf{Q}_{(1)} + ... \label{bbQn}
\end{equation}
where
\begin{equation} \mathbb{P}_{(n)} = \int m
{\bf c} {\bf c} f_{(n)} d{\bf v} \qquad {\rm and} \qquad {\bf
Q}_{(n)} = \int {1\over 2} m {\bf c^2} {\bf c} f_{(n)} d{\bf v}\ .
\label{fluxion} \end{equation}
Then we readily see that \begin{equation} \mathbb{P}_{(0)} =
\int_0^\infty m
c^4 f_0 dc \int {\bf e}\, {\bf e}\, d\Omega \label{P0}
\end{equation} where $c=|{\bf c}|$ and ${\bf e}={\bf c}/c$.

With a little rearranging, the first of the two integrals in (\ref{P0})
becomes proportional to a gamma function. The second can be written as
\begin{equation} \int {\bf e}\, {\bf e}\, d\Omega = {4\pi \over 3}
\mathbb{I} \label{I}
\end{equation} where $\mathbb{I}$ is the unit tensor; its components
are those of the Kronecker delta.  Hence, we find that \begin{equation}
\mathbb{P}_{(0)} = p\, \mathbb{I} \label{pressT} \end{equation}
where \begin{equation} p = R \rho T \ , \label{press}
\end{equation} which we identify with the gas pressure.  By
contrast, the integral for the heat flux is odd in ${\bf c}$ and
we find that ${\bf Q}_{(0)}={\bf 0}$.  In short, the macroscopic
equations at leading order are the Euler equations.

To evaluate $\mathbb{P}_{(1)}$, we note that \begin{equation}
\mathcal{D} = {D\over Dt} + {\bf c}\cdot \nabla + {\bf a}\cdot
\partial_{\bf c} \label{calD} \end{equation}
where \begin{equation} {D\over Dt} = \partial_t + {\bf u}\cdot
\nabla \label{DDt} \end{equation} and we use $\nabla$ and
$\partial_{\bf x}$ interchangeably. Then we have
\begin{equation} f_{(1)} = - f_0 \left[\mathcal{A} + {\bf c}\cdot
{\bf B} + {c^2\over 2RT} {D \over Dt} \ln T - ({\bf cc}:\nabla \ln
T)/(RT) + {c^2\over 2RT} {\bf c}\cdot \nabla \ln T \right]
\label{f1ex} \end{equation} where \begin{equation}
\mathcal{A} = {D \over Dt} \ln {\rho
\over T^{3/2}}, \quad {\bf B} = \nabla \ln {\rho \over T^{3/2}} +
\left({\bf a} - {D{\bf u} \over Dt}\right)/(RT) \ . \label{AB}
\end{equation}

With the ${\bf c}$-related factors thus in evidence, it is a
straightforward matter to carry out the necessary integrals to
evaluate $\mathbb{P}_{(1)}$ and ${\bf Q}_{(1)}$.  For this, we need the
formula \begin{equation}
\int e^i\,e^j\,e^k\,e^\ell\, d\Omega = {4\pi \over 15} \left(\delta^{ij}
\delta^{k\ell} + \delta^{ik} \delta^{j\ell} + \delta^{i\ell}
\delta^{jk}\right) \; ,\label{4pi}
\end{equation}
which may be verified by explicit evaluation of its components.
If we include $\mathbb{P}_{(0)}$,  we may write the result of the
integration as \begin{equation}
\mathbb{P} = \left[p - \mu\left({D \ln T \over Dt} + {2\over 3} \nabla
\cdot {\bf u} \right)\right]\, \mathbb{I} - \mu \, \mathbb{E}
+ \mathcal{O}(\tau^2) \label{x39} \end{equation} where $\mu=\tau p$, and
\begin{equation}
E^{ij} = {\partial u^i\over \partial x_j } + {\partial u^j\over
\partial x_i } - {2\over 3} \nabla \cdot {\bf u} \, \delta^{ij} \
. \label{E} \end{equation}
For the heat current, we get
\begin{equation} {\bf Q}= -\eta\nabla T  - \eta T \nabla\ln p -
{5\over 2}\mu \left({D{\bf u} \over Dt} - {\bf a}\right) +
\mathcal{O}(\tau^2) \label{Qform} \end{equation} where
$\eta=\frac{5}{2}\mu R$.

A more detailed derivation is given in~\cite{che00c}.  The expressions
for $\mathbb{P}$ and ${\bf Q}$ involve not only the
fluid dynamical fields, or lower moments of $f$, but their substantial
derivatives as well.  Those derivatives are given by the fluid dynamical
equations that, in turn, involve $\mathbb{P}$ and ${\bf Q}$. So the higher
moments are given only implicitly by (\ref{E}) and (\ref{Qform}) and we
need to solve the fluid equations together with those equations to obtain
explicit expressions for the higher moments.

\section{Discussion of the Equations}

\subsection{The Entrance of Entropy}

Since the equations for $\mathbb{P}$ and ${\bf Q}$ are
intertwined with the field equations themselves, we
first rewrite these equations so as to clarify their
meaning.  We may introduce the continuity equation into (\ref{E})
and so replace $\nabla \cdot {\bf u}$ by $-\dot \rho/\rho$ where the
dot stands for $D/Dt$.  The combination of $\dot T/T$ and of $\dot
\rho/\rho$ that then appears suggests the introduction of the quantity
\begin{equation}
S = {3\over 2} R \ln {p\over \rho^{5/3}} . \label{entropy}
\end{equation}
For an ideal gas, this is the formula for the specific entropy, with
$C_v=3R/2$ and $\gamma=5/3$.  If we introduce $S$ together with the
definition of $p$, we may rewrite (\ref{x39}) as \begin{equation}
\mathbb{P} = p\left[1 - {\tau \dot S\over C_v}\right]
\, \mathbb{I} - \mu \mathbb{E}
+ \mathcal{O}(\tau^2) \ . \label{xx39} \end{equation}

Similarly, the momentum equation (\ref{mom}) can be used to rewrite the
formula for the heat flux.  Since $(\ref{Qform})$ contains $D{\bf u}/Dt$,
we may use (\ref{mom}) to rewrite it as \begin{equation}
\mathbf{Q} = -\eta \nabla T - {5\over 2} \nu \nabla \cdot
\left({\tau p \dot S\over C_v}\, \mathbb{I} + \tau p \mathbb{E}\right)
+ \mathcal{O}(\tau^2) \ ,
\label{newQ} \end{equation}
where $\nu=\mu/\rho$.

We may also use (\ref{entropy}) to convert (\ref{heat}) into an
evolution equation for the specific entropy.  If we introduce
(\ref{xx39}) into that equation, we find a term
$p \nabla \cdot {\bf u}$ so that $\dot \rho/\rho$ comes in by way
of the continuity equation.  The term in (\ref{heat}) involving $\dot
T/T$ combines with this and we obtain an equation for $\dot S$. But
$\dot S$ also appears in $\mathbb{P}$ and hence, when we gather the two
apparitions of $\dot S$ together, we see that (\ref{heat}) becomes
\begin{equation} \rho T (1-{2\over 3} \tau
\nabla \cdot {\bf u}) {DS \over Dt} = -\mu \mathbb{E}:\nabla {\bf u} -
\nabla \cdot {\bf Q}  \ . \label{entro}
\end{equation}

Though our equations may have an unfamiliar look, this is not because we
have done anything unusual.  Rather, we have omitted to do some things
that are normally considered usual.  So let us see how
to get back to more familiar ground.

\subsection{The Euler and Navier-Stokes Equations}

We have obtained \begin{equation}
\mu = \tau p \qquad {\rm and} \qquad \eta = {5\over 2} \tau p R
\label{transco} \end{equation}
and so we see that when $\tau \rightarrow 0$, we have $\mathbb{P}=
p\,\mathbb{I}+\mathcal{O}(\tau)$ and ${\bf
Q}=\mathcal{O}(\tau)$, with $\dot S =\mathcal{O}(\tau)$ according
to (\ref{entro}). Thus we obtain the Euler equations in leading order
together with entropy conservation.

We see also that the term $\tau \dot S/C_v$ in both (\ref{xx39}) and
(\ref{newQ}) is of order $\tau^2$.  Therefore, we find that
\begin{equation}
\mathbb{P}=
p\,\mathbb{I} - \mu\, \mathbb{E} + \mathcal{O}(\tau^2)
\label{firstP} \end{equation} and \begin{equation} {\bf Q}= - \eta
\nabla T + \mathcal{O}(\tau^2),\end{equation}
when $\tau$ is very small.
Therefore, when the extra terms in our pressure tensor and heat flux
are developed in $\tau$, we see that our forms differ from the standard
Navier-Stokes terms in terms of order $\tau^2$, which is perfectly
allowable in first-order theories.

In contrast to the conventional closure approximations, our expressions
for $\mathbb{P}$ and ${\bf Q}$ depend on both the fluid fields and their
derivatives.  This means that these expressions must be solved in concert
with the dynamical equations.  To express $\mathbb{P}$ and ${\bf Q}$
explicitly in terms of the fluid fields, as in the usual closure
relations, we would need to make expansions in $\tau$.  As we have just
seen, in the first order, we recover the Euler equations and in the second
order we get the Navier-Stokes equations.  Continuation of this
development produces terms of all orders in $\tau$.  Therein lies the
crucial  difference of our results from those of the Chapman-Enskog
procedures.  The extra terms in our development of the present results do
not correspond to the higher theories based on Chapman-Enskog procedures
that lead to the Burnett equations, as shall be explained in another
place.  All the terms in the present approximation come about from a
first-order theory and the differences from standard theory arise
in terms of second and higher order.  When  $\tau$ is not
infinitesimal, which it never is in practice, these terms do have an
effect on the predictions of the theory.

\subsection{Dynamical Pressure}

To clarify the meaning of the difference between our equations for a
simple gas and those obtained with the Chapman-Enskog procedure
we note that, as in \ref{Hilb}, we are writing the
solution of the kinetic equation at any order as \begin{equation}
f = f_N + \mathcal{O}(\tau^{N+1}) \qquad {\rm with} \qquad f_N =
\sum_{m=0}^{N} f_{(m)} {\tau}^m  \ . \label{approx} \end{equation}
From this we then obtain an approximation for the stress tensor
in the form \begin{equation}
\mathbb{P} = \mathbb{P}_N + \mathbb{R}_N  \label{err} \end{equation}
where $\mathbb{R}_N = \mathcal{O}(\tau^{N+1})$ is the error incurred
in the truncation of the series.  If we take the trace of (\ref{err}),
we find \begin{equation}
{\rm tr}\; \mathbb{P} = {\rm tr}\; \mathbb{P}_N + {\rm tr}\;
\mathbb{R}_N \label{trace}\ . \end{equation}

As we see from the definition (\ref{ptensor}) of $\mathbb{P}$, in the
exact case,
\begin{equation} {\rm tr}\,\mathbb{P} = 3p . \end{equation}
This result says that the total pressure has
no contribution of a dynamical kind for a structureless gas,
as Maxwell and Boltzmann both recognized.  The Chapman-Enskog
procedure imposes this condition at every finite order so that
\begin{equation}
{\rm tr}\; {\mathbb{P}}_N^{CE}=3p \ . \label{COND} \end{equation}
When we impose (\ref{COND}) onto (\ref{trace}), we are forcing the
requirement that ${\rm tr}\; \mathbb{R}_N=0$, which overly
constrains the results obtained by Chapman and Enskog and
destroys any hope of improving
convergence by mitigating their errors.  However, all that we should
demand of our successive approximations is that in the $N$th
approximation,
\begin{equation}
{\rm tr}\,\mathbb{P}_N = 3p + \mathcal{O}(\tau^{N+1})
\end{equation} On introducing this less restrictive condition
into (\ref{trace}) we find that, in our procedure,
\begin{equation} {\rm tr}\;
\mathbb{R}_N = \mathcal{O}(\tau^{N+1})\ , \end{equation} which
does not incur the loss of generality that forcing the trace of
$\mathbb{R}_N$ to vanish does. By keeping terms
$\mathcal{O}(\tau^{N+1})$
in our $N$th approximation for $\mathbb{P}$ and ${\bf Q}$, we leave
open the possibility of compensating for the errors caused by the
truncation of the series for $f$ by retaining
suitable process-dependent effects of higher order.

Thus we have in our present approximation a dynamical pressure in
our approximate pressure tensor, $\mathbb{P}_1 =
\mathbb{P}_{(0)} + \tau \mathbb{P}_{(1)}$, that is,
\begin{equation}
{\rm tr}\; \mathbb{P}_1 = 3p\left[1-{\tau \dot S\over C_v}\right]
= 3p + \mathcal{O}(\tau^{2})\ .  \end{equation}
Our asymptotics suggests that the extra term compensates for the
effect of truncation of the series for $f$.  Since this term has
contributions from all orders in $\tau$, it can in principle be
very effective in extending the domain of validity of the theory.

\section{Conclusion}

We have illustrated our derivation of the fluid equations from kinetic
equations by carrying out the procedure for the relaxation model of
kinetic theory.  The same procedure can be used on other forms of the
kinetic equation.  In the case of the Boltzmann equation, the procedure is
quite similar, though the inversion of the linearized Boltzmann collision
operator involves some technical issues that we shall take up elsewhere.
However, by working out the case of the relaxation model we can more
readily see the differences between our approach and the Chapman-Enskog
procedure.

In our derivation, we do not introduce slow times as in the Chapman-Enskog
method.  This means that we are not driven to expand the fluid variables
(or slow quantities) in $\tau$ as in C-E theory.  Such expansions cause
ambiguity in the application of initial conditions since it is not clear
how to distribute the initial values over the various orders.  Moreover,
those expansions lead to a different sequence of approximations than ours.
Chapman-Enskog theory gives the pressure tensor and the heat flux
explicitly in terms of the fluid fields, a feature which results from
invoking solvability conditions at each order.  Our results
do not produce explicit formulae for $\mathbb{P}$ and ${\bf Q}$ in terms
of the fluid fields; rather, these quantities are expressed in terms of
the fluid fields {\it and} their derivatives.  Those derivatives appear
in the field equations themselves so that we do produce a closed system of
equations.  Moreover, since the expansion variable $\tau$ appears in the
equations, we may further expand the equations to develop explicit
formulas for $\mathbb{P}$ and ${\bf Q}$ accurate to any prescribed order.

As we saw, the leading terms in the development of the formula for
$\mathbb{P}$ and ${\bf Q}$ give us successively the Euler and
Navier-Stokes equations, but
the development need not stop there.  That is, our finite formulae
implicitly contain terms of all orders in $\tau$ and this means that we
may hope that they will produce high accuracy even when $\tau$ is not
infinitesimal.  In any event, it is clear that we may expect a divergence
between results from our system and those from the Navier-Stokes equations
when $\tau$ is not very small.

J.B. Keller (private communication) has remarked that two theories with
the same nominal accuracy may have different domains of
validity.  In forcing the trace of $\mathbb{P}_N$ to be exactly $3p$,
order by order, the Chapman-Enskog method renounces the extra generality
allowed by the freedom to choose higher order terms in an advantageous
way.  For us, the problem has been to select the best way to allow for the
higher order corrections when trying to extend the domain of validity of
the theory.  We shall use as our test of validity the comparison with
experiment given in the two following papers of this series.

\goodbreak

\end{document}